\documentclass[reprint, 
superscriptaddress,
 amsmath,amssymb,
 aps,
]{revtex4-1}

\usepackage{graphicx}
\usepackage{dcolumn}
\usepackage{bm}
\usepackage{color, gensymb}

\usepackage{caption} 

\def\psik {\Psi_i \rangle}
\def\psib {\langle \Psi_i}
\def\psnk {\Psi_n \rangle}
\def\psnb {\langle \Psi_n}

\def \dH {\frac{\partial H}{\partial R_\alpha}}
\def\pa {\partial}

\def \asi {{\it a}-Si}
\def \lsi {{\it l}-Si}

\begin{document}

\preprint{APS/123-QED}

\title{Sculpting the band gap: a computational approach}
\author{Kiran Prasai}
\affiliation{%
Department of Physics and Astronomy, Clippinger Laboratories\\
Ohio University, Athens, OH 45701
}%
\author{Parthapratim Biswas}
\affiliation{%
Department of Physics and Astronomy \\
The University of Southern Mississippi, Hattiesburg, MS 39406
}%
\author{D. A. Drabold}%
\affiliation{%
Department of Physics and Astronomy, Clippinger Laboratories\\
Ohio University, Athens, OH 45701
}%
\date{\today}
%
\begin{abstract}
Materials with optimized band gap are needed in many specialized applications.
In this work, we demonstrate that Hellmann-Feynman forces associated with the gap states can
be used to find atomic coordinates with a desired electronic density of states.
Using tight-binding models, we show that this approach can be used to arrive at 
electronically designed models of amorphous silicon and carbon. We provide a 
simple recipe to include {\it a priori} electronic information in the formation
of computer models of materials, and prove that this information may have
profound structural consequences. An additional example of a graphene 
nanoribbon is provided to demonstrate the applicability of this approach 
to engineer 2-dimensional materials.  The models are validated with plane-wave density functional 
calculations.\end{abstract}

\pacs{Valid PACS appear here}
\maketitle


The central goal of materials science is the development of materials with novel properties. In general, this program of materials engineering has proceeded largely by experimental exploration.  In this Letter, we offer a novel and direct approach to determining structures (e.g. atomic coordinates) that yield desired electronic or optical properties. The method is ÒdirectÓ in that an initial structure is purposefully modified to push the model toward a desired electronic density of states (for example, engineering a gap of desired magnitude, eliminating defect states in the gap, or perhaps changing the structure of band tails, a serious issue in some 
photo voltaic (PV) applications \cite{liang1}. Such a tool is especially 
valuable for semiconductors, and may have value for chalcogenide phase change 
memory materials and device applications.  
The outcome of these simulations is a set of coordinates that satisfy desired optical properties, within the accuracy of the Hamiltonian employed. Naturally, there is no {\it a priori } 
guarantee that a model with preferred electronic properties should be a minimum of the total energy functional, but we show in two examples how to find electronically optimized configurations that are indeed stable, and that the method is a powerful aid to finding realistic models of {\it a}-Si 
using an electronically modified liquid-quench approach (normally slow and impractical, 
but effective in our approach). We provide a means to obtain relaxed models of {\it a}-C 
with widely varied sp$^2$/sp$^3$ ratios, and employ the method to show 
how to open up a large optical gap in a graphene nanoribbon. 
Models are validated with plane-wave density 
functional theory (DFT)\cite{kresse1,kresse2}. 

To introduce our method, we adopt tight-binding Hamiltonians and employ 
Hellmann-Feynman forces~\cite{martin, hft} in a novel way to determine structures 
with desired optical gap. Recall that the spatially non-local part of the 
interatomic force, the band-structure force, has the form:
\begin{equation}\label{eq_a}
\Vec{F}^{BS}_{\alpha} =  -\sum^{occ} \limits_{i} \langle \Psi_{i}(r)| \frac{\partial H}{\partial R_{\alpha}}|\Psi_{i}(r) \rangle 
\equiv \sum^{occ} \limits_{i} \Vec{F}^{BS}_{i,\alpha}. 
\end{equation}

\noindent Here $ i $ indexes eigenvalues (or bands), $R_\alpha$ are the 3$N$ positional degrees of freedom, $H$ is the Hamiltonian, and $\Psi_i$ is an eigenvector. For a complete basis set (such 
as plane waves), the contributions from the nuclear derivatives of the wave function in the first two terms cancel exactly (the Hellmann-Feynman theorem~\cite{martin, hft}).
If one considers individual terms in the sum in Eq.(\ref{eq_a}), the term $\Vec{F}^{BS}_{i}$ represents the contribution from the $i^{th}$ band (or eigenvalue) to the total bandstructure force. In effect, $\Vec{F}^{BS}_{i}$ is a gradient for the $ i^{th} $ energy eigenvalue $ \lambda_{i} $. 
As such, $\Vec{F}^{BS}_{i}$ provides the direction in the 3N-dimensional 
configuration space of most rapid change of  $\lambda_{i}$.  Thus, to shift  $ \lambda_{i} $ to higher (lower) energies, we should move atoms incrementally along the direction --$\Vec{F}^{BS}_{i} 
\text{ (}+ \Vec{F}^{BS}_{i} \text{)}$. So, we simply move the atoms along (or against) 
these gradients to push tail/defect states away from the gap into the valence or conduction band. For small displacements  $\delta R_{\alpha}$ along this gradient, the shift $\delta \lambda_i$ of an eigenvalue $\lambda_i$ can be written as,
\begin{equation}\label{eq_b}
\delta \lambda_{i} = \sum \limits_{\alpha} \frac{ \partial \lambda_{i}}
{\partial R_{\alpha}} \delta R_{\alpha} = \sum \limits_{\alpha} 
- \Vec{F}^{BS}_{i,\alpha} ~ \delta R_{\alpha}. 
\end{equation}
\noindent To this end, we introduce the term {\it{gap force}} for state $i$ to indicate the force (negative nuclear gradient) associated with eigenvalue $\lambda_i$. We exploit such forces to push eigenvalues out of a spectral range that we wish to be free of states. In semiconductors or insulators,  the gap region may be sparsely populated with defect and impurity states, possibly localized,  in which case the gap forces can be viewed as local perturbations arising from an effective energy 
functional~\cite{atta}, 
\begin{eqnarray}
\Phi(R_1, R_2 \ldots, R_{3N}) & = & \sum_i f_i \psib | H | \psik + U_r \nonumber \\
&+& \sum_n^{\prime} \gamma g(\lambda_n) \left(\psnb |H|\psnk - \varepsilon_f
\right)
\label{eq_d}
\end{eqnarray}
\noindent The sum in the last term in eq.\,(\ref{eq_d}) is restricted to gap states we wish to clear in a spectral range $[E_{min}, E_{max}]$. Here,  $g(\lambda_n)$ = +1 or -1 
for $\lambda_n > \lambda_{HOMO}$ or $\lambda_n \le \lambda_{HOMO}$ respectively, 
and $f_i$ is the occupation number of $i^{th}$ energy level, which is either 0, 1, or 2. The parameter $\gamma$ controls the strength of the gap force, $\varepsilon_f$ is the Fermi energy, and $U_r$ is the repulsive ion-ion interaction. The force associated with the $\alpha^{th}$ degree of freedom is given by, 
\begin{equation}
\Vec{F}_{\alpha}^{bias} = \Vec{F}_{\alpha}^{BS} + \Vec{F}_{\alpha}^{ion} 
+ \Vec{F}_{\alpha}^{gap}, 
\label{eq_e}
\end{equation}
\noindent which can be used to obtain strong local minima by minimizing the total energy and forces via MD simulations and/or relaxations. In the tight-binding formulation, the forces on the 
right-hand side of (Eq.\,\ref{eq_e}) are:
\begin{eqnarray} 
\Vec{F}_{\alpha}^{BS} 
& = &   -\sum \limits_{i} f_i \langle \Psi_{i}(r)| \frac{\partial H}{\partial R_{\alpha}}|\Psi_{i}(r) \rangle \nonumber \\
\Vec{F}_{\alpha}^{ion} & = & -\frac{\pa U_r}{\pa R_{\alpha}} \nonumber \\
\Vec{F}_{\alpha}^{gap} & = & -\sum^{E_{max}}_{i=E_{min}} \gamma g(\lambda_i) 
\psib | \dH | \psik \nonumber
\end{eqnarray} 
We show empirically that the method works well even for midgap states near $\varepsilon_{F}$.  
We have observed that the method is also applicable in the opposite mode: to {\it maximize} 
the density of states at the Fermi level by shepherding eigenvalues toward the 
Fermi level \cite{unp1}. This might, for example, introduce new structural features 
and produce models with interesting electrical conductivity.

In the rest of the paper, the first two examples provide new equilibrium (i.e. relaxed) 
models of {\it a}-Si and {\it a}-C using biased dynamics to obtain a desired gap. 
For these, we perform melt-quench simulations with biased forces as in Eq.~\ref{eq_e} and always take $\gamma=1$. Since we seek models at a minimum of the unmodified tight-binding total-energy functional (Eq.~\ref{eq_d} with $\gamma=0$), we then relax the system by damping the velocity of atoms until the forces on atoms vanish. This simplest prescription has proven adequate to create with the additional flexibility of ``optical sculpting". The third example, tersely described, reports a structure of a graphene nanoribbon without concern for making physical forces vanish.

We undertake our first calculation on {\asi}. It is known for {\asi},  that rapid melt-quenching from the liquid produces many gap states, remnant of the~6-fold liquid metal and an unsatisfactory model for a tetrahedral amorphous semiconductor. In this example, we impose biased dynamics favoring the creation of a gap in the range observed in the best available WWW models~\cite{wooten}. After equilibrating the liquid in the conventional way, we quench with dissipative dynamics (velocity rescaling to 
300 K) and biased forces. We note, for the choice of $\gamma=1$ we invoke, that the average gap forces remain less than 20\% of the TBMD forces on atoms: the dynamics is modified in a 
somewhat subtle way, but operating over many steps, the method 
yields structures improved both optically (by construction) 
but also {\it structurally}, a considerable bonus and proof 
that inclusion of electronic {\it a priori} information does 
influence structure, and in a way that improves agreement 
with experiments. We use the Goodwin-Skinner-Pettifor (GSP) 
Hamiltonian for Si \cite{goodwin1}.

In the quenching process, we add gap forces to push the 
eigenvalues away from the Fermi level, using 
prescription \ref{eq_e}.  Since we relax the model with $ \gamma =0$
after quenching, the final model is at equilibrium 
according to the {\it true} forces derived from 
\ref {eq_d} with $\gamma=0$. 

\begin{figure}
\includegraphics[width=0.5\textwidth]{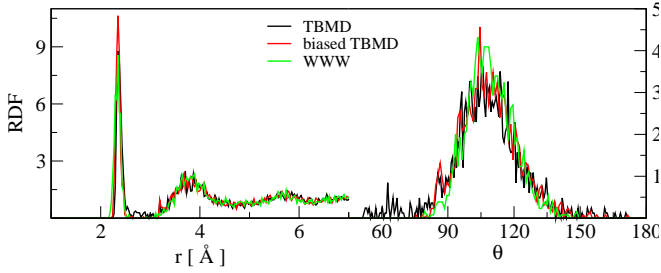}
\caption{\label{rdf1}
{\small (Color online)
Radial (left) and bond-angle (right) distributions of {\asi} from three 
models: biased TBMD (red), TBMD (black), and WWW (green)~\cite{wooten}.
}
}
\end{figure}

\begin{figure}
\includegraphics[width=0.4\textwidth]{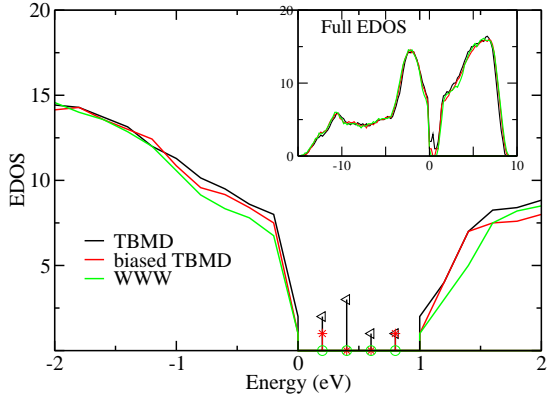}\\~\\ 
\caption{\label{dos} 
{\small (Color online) 
Density of electronic  states (EDOS) near the band-gap 
region for three models of {\asi}: TBMD, biased TBMD 
and WWW\cite{wooten}. The density of states within the gap 
is shown as a vertical line with respective color as 
indicated. Inset shows full EDOS. 
}
}
\end{figure}

A 216-atom model of liquid silicon ({\lsi}) was prepared 
by cooling an initial random configuration from a
temperature of 2500 K to the melting point $\approx$
1780 K in several steps, which were followed by
equilibration for a period of 50 ps per step and 
total-energy relaxations.  We have verified that our {\lsi} model produces 
features similar to {\lsi} models obtained from the GSP Hamiltonian by earlier 
workers~\cite{wang1, kim1, kwon1}. In figs.\,\ref{rdf1}-\ref{dos}, we present
structural and electronic properties from the dynamics for $\gamma=0$ (e.g. conventional MD) 
and 1 starting with liquid silicon.

Figure \ref{rdf1} compares the radial and bond-angle distribution 
functions for three models of {\asi} obtained 
from conventional TBMD, biased TBMD, and the 
WWW method~\cite{wooten}, the last being the ``gold standard" of {\it a}-Si 
models.  A comparison of the 
RDFs and an analysis of bond lengths shows that 
the biased dynamics significantly increases the short-ranged order: 
the biased model exhibits fewer short and long bonds as 
compared to conventional TBMD and WWW models. 
A remarkable feature of the biased-MD model 
is the absence of 60{\degree} bond angles.  These 
angles are typically associated with frozen liquid-like 
configurations, which plague conventional 
MD simulations by producing 3-fold 
defects (dangling bonds). The latter are notoriously difficult 
to remove from MD models via conventional means and 
have detrimental effects on electronic and optical 
properties. 
It is noteworthy that the inclusion of gap forces 
can eliminate these unphysical features 
completely and impart more tetrahedral order 
in the structure. Atomic coordination is also 
markedly improved in biased dynamics. 
Atomic-coordination statistics show that 
97.2\% atoms in the biased-MD model are 4-fold 
coordinated.  This result is not only superior 
to 87\% 4-fold coordination in conventional MD 
but also better than earlier works reported in the 
literature using the GSP Hamiltonian~\cite{wang1, kim1, colombo}. 
The biased-TBMD model has fewer defects around the 
Fermi-energy than the TBMD model (fig. \ref{dos}). 
Next, we check these models using using DFT in 
local density approximation (LDA). 
LDA calculations with VASP show that the energy of the 
biased-TBMD configuration is lower than that of the conventional 
TBMD configurations. {\it Imposing electronic
constraints leads to relaxed models in better 
agreement with structural experiments.}

To explore the reproducibility of our method, we 
sampled 25 well-separated {\it l-}Si configurations
as starting coordinates and quenched these configurations 
with $ \gamma = 0 \text{ and } 1$ as described earlier. 
This was followed by total-energy relaxations of 
the models to their respective local minimum. 
We analyzed these models to obtain the number 
of atoms with 4-fold coordination as a measure of 
the merit of the model. Of 25 biased-TBMD models that 
we studied in this work, 20 models were found to 
produce better (more tetrahedral) topology than
TBMD models.

As a second example we study {\it a}-C. Tetrahedral amorphous carbon ({\it ta}-C), 
has some properties reminiscent of diamond while 
potentially holding some advantages\cite{robertson, drabold2}. 
The tight-binding model of Xu et al \cite{xu1} has been used 
previously to model {\it ta}-C with limited success \cite{wang2}. 
These calculations involved a quench from a high-density liquid ({\it l}-C) 
and volume rescaling at lower temperature. Using the same Hamiltonian, 
we demonstrate that a simpler melt-quench method can yield improved models.
Amorphous carbon dominated by $sp^3$ bonding is characterized by 
a band gap of about 2 eV (depending on the fraction 
of $sp^{3}$ bonded atoms) in contrast to $sp^{2}$-bonded 
{\it a}-C,  which has a gap of less than 0.5 eV \cite{robertson}. 
The perfectly $sp^{3}$-bonded WWW model of {\it ta}-C \cite{wooten}, 
relaxed with the Xu Hamiltonian, has a gap of 4.1 eV.  
We used this spectral range with biased melt-quench TBMD to form 
models without states in the gap region as exhibited by WWW 
models.

Starting with liquid carbon of density 3.5 $gm/cm^{3}$ 
equilibrated at 10000 K, we quenched the model to 700 K 
at a rate of 500 K/ps. At this point, two parallel runs 
(quenching) were performed: one with regular TBMD forces 
($\gamma=0$) and the other with additional gap forces for 
$\gamma=1$. Following first example, we designate these runs as 
`TBMD' and `biased TBMD', respectively. Both quenched 
models were relaxed to their respective local minimum 
until the  force on each atom is less than 0.05 eV/{\AA}. 
Gap forces were switched off during post-quench relaxation for 
biased-TBMD models. The magnitude of gap forces 
were found to be smaller ($< 20\%$) than the 
corresponding TBMD forces throughout the quenching 
process.

The unaided TBMD with the Xu Hamiltonian prefers $sp^{2}$ 
dominated network as observed in our calculations and 
in Refs. \cite{xu1, wang4}. The diamond-like $sp^{3}$-bonded networks 
reported in \cite{wang2} appear to be an artifact of high density or 
high pressure on {\it l}-C. Our calculations produced models 
with up to 94\% 4-fold coordination compared with 74\% and 89\% in \cite{wang2}. 
Our results are readily reproducible and do not involve arbitrary 
manipulation of density. We have conducted 25 quenching runs 
with different starting liquid models and all of these models 
produce tetrahedral networks with more than 90\% 4-fold 
coordination. 
\begin{figure}
\includegraphics[width=0.5\textwidth]{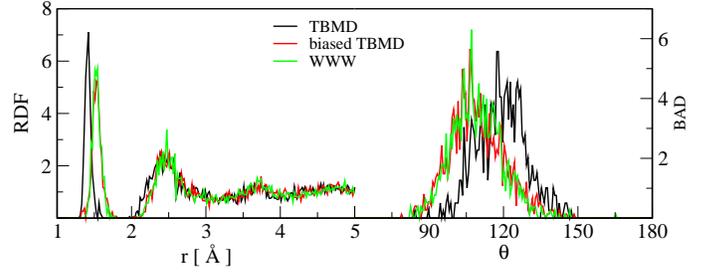}
\caption{\label{adim_str}
{\small (Color online)
Radial (left) and bond-angle (right) distribution functions 
of {\it ta}-C from biased TBMD, TBMD, and WWW. See text for 
discussion of results.}
}
\end{figure}
\begin{figure}
\includegraphics[width=0.5\textwidth]{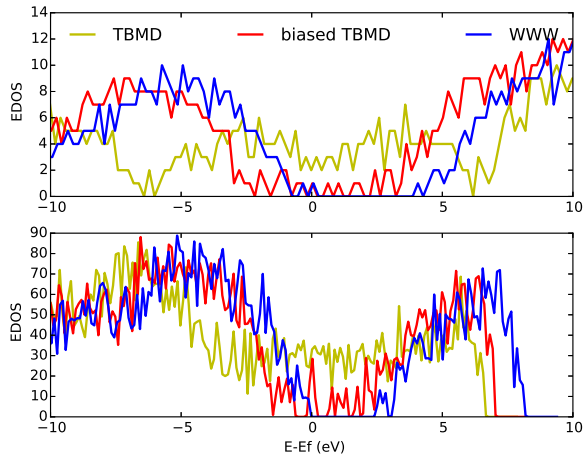} 
\caption{\label{dos_C}
{\small (Color online)
Density of electronic  states (EDOS) for three models of {\it a-}C: TBMD, biased TBMD 
and WWW\cite{wooten}. The upper plot is obtained using the tight-binding model of \cite{xu1}
and the lower one from LDA \cite{kresse1,kresse2}}
}
\end{figure}
The structural features, including RDF, BDF and atomic coordinations from 
biased TBMD resemble closely with {\it ta}-C WWW model. The TBMD model 
is dominated by $sp^{2}$ bonding and registers distinct peaks in the RDF 
and BDF (Figure \ref{adim_str}). The density of electronic states of the biased-TBMD model 
shows that the gap opens up to 0.7 eV as compared to 0.21 eV 
in the TBMD model. Also, the biased-TBMD model has only 14 states 
in the gap region exhibited by WWW models as compared to 71 states 
in the TBMD model: see Fig. \ref{dos_C}. The electronic structure of these models is 
also confirmed by LDA calculations using VASP. The biased-TBMD 
model shows few scattered states in the gap region as opposed to 
the `metal-like' electronic structure of the TBMD model. Total 
energy calculations using LDA show that the biased-TBMD model 
has 0.31 eV/atom less energy than the regular TBMD model. The biased 
model is also stable under relaxation using LDA. Such relaxation 
decreases the total energy by 0.07 eV/atom while preserving 
structural ordering of the model. 

Finally, we have applied  this approach to graphene nanoribbons. 
Using a $\Gamma$-point calculation with the same TB Hamiltonian 
in example 2 on an armchair graphene nano-ribbon (AGNR) of 
width 11.07 {\AA}, we have carried out low-temperature MD simulations 
biased toward an enlarged bandgap, without trying to drive the model 
to a minimum of the total energy. Depending on the value of $\gamma$ 
selected, we obtain structures with gaps of up to 1.58 eV (wider 
gap is observed using LDA). But these computer models have high strain 
on the edge atoms and the length of C-C dimers along the edge 
becomes smaller than the average bond length by 18\%.

As we have shown, our method is best employed in a 
``statistical mode"--unsurprisingly the final structures 
depend on the initial state. In some fraction ($ \approx 20\%$) 
the method does not improve the gap in case of {\it a-}Si. We 
suppose that this may be due to the very simple rule of shifting atoms along gradients
toward the nearest band edge, even for eigenvalues very near $ \varepsilon_{f} $.

For  {\it a}-Si, we use {\it a priori} knowledge of the gap from 
the best available models.  In the general case, one can define a gap 
by trial and error, with the choice being determined in part by 
a requirement that the physical forces vanish at the end. For 
{\it a}-C, considerable flexibility is afforded by our approach 
in tuning $sp^2/sp^3$ ratios. We have experimented with 
various $\gamma$, and have found no particular advantage to 
selecting $\gamma \neq 1$ to date.  Furthermore, preliminary studies 
suggest that the results presented here also accrue for larger 
(512-atom) models. We expect that the scheme will be useful for 
many other complex materials not only for discovering structures 
with desired gaps but also for imposing electronic constraints 
in modeling.

As is the case with all methods, our approach has limitations: (1) For this first report we use standard tight-binding Hamiltonians for the simulations. Such Hamiltonians are well known to have imperfect transferability (for this reason we are currently extending the scheme to plane-wave DFT, a straightforward but tedious undertaking) and 2) even in a density-functional framework, gap estimates from Kohn-Sham eigenvalues are spurious, though usually these account reasonably well for trends. With significant computational expense, these estimates may be improved {\it e.g.} with GW or Hybrid Functional schemes \cite{martin}. Despite these limitations, we demonstrate the utility of the method with two examples and suggest that the approach may be developed in promising ways.  Nevertheless, it is plain that the method is an extremely useful new tool, even using this simplest implementation.
\begin{acknowledgments}
DAD thanks Army Research Office for supporting this work 
under Grant W911NF1110358, and Ohio Supercomputer Center. 
PB acknowledges 
support from the International Materials 
Institute for New Functionality in Glass via NSF grant 
no. DMR-0844014.
\end{acknowledgments}
%
%
%
\nocite{*}
\bibliography{kbib}
\end{document}